\pdfoutput=1

\documentclass[11pt,twoside,a4paper,cmspaper,final,collab]{cms-tdr}
\def\svnVersion{175803}\def\cmsCernNo{2012-329}\def\cmsCernDate{\today}\def\cmsMessage{Submitted to Physical Review D}

\begin{document}\cmsNoteHeader{EXO-11-009}

\hyphenation{had-ron-i-za-tion}
\hyphenation{cal-or-i-me-ter}
\hyphenation{de-vices}

\RCS$Revision: 162004 $
\RCS$HeadURL: svn+ssh://svn.cern.ch/reps/tdr2/papers/EXO-11-009/trunk/EXO-11-009.tex $
\RCS$Id: EXO-11-009.tex 162004 2012-12-19 02:33:59Z alverson $

\ifthenelse{\boolean{cms@external}}{%
\newcommand{\scotchrule[1]}{\centering\begin{ruledtabular}\begin{tabular}{#1}}
\newcommand{\donescotchrule}{\end{tabular}\end{ruledtabular}}
}{
\newcommand{\scotchrule[1]}{\centering\begin{tabular}{#1}\hline\hline}
\newcommand{\donescotchrule}{\hline\hline\end{tabular}}
}
\newlength\cmsFigWidth
\ifthenelse{\boolean{cms@external}}{\setlength\cmsFigWidth{0.5\textwidth}}{\setlength\cmsFigWidth{0.7\textwidth}}

\cmsNoteHeader{EXO-11-009}

\title{\texorpdfstring{Search for contact interactions in $\mu^+\mu^-$ events in $\Pp\Pp$
collisions at $\sqrt{s} = 7\TeV$}{Search for contact interactions in opposite-sign dimuon events in pp
collisions at sqrt(s) = 7 TeV}}

\address[wsu] {Wayne State University}

\author[wsu] {S. Gollapinni}
\author[wsu] {P. Karchin}
\author[wsu] {C. Kottachchi}
\author[wsu] {P. Lamichhane}
\author[wsu] {M. Mattson}
\author[wsu] {C. Milst\`ene}

\date{\today}

\abstract{Results are reported from a search for the effects of
  contact interactions using events with a high-mass,
  oppositely-charged muon pair. The events are collected in
  proton-proton collisions at $\sqrt{s} = 7$\TeV using the Compact
  Muon Solenoid detector at the Large Hadron Collider. The data sample
  corresponds to an integrated luminosity of 5.3\fbinv. The observed
  dimuon mass spectrum is consistent with that expected from the
  standard model.  The data are interpreted in the context of a quark-
  and muon-compositeness model with a left-handed isoscalar current
  and an energy scale parameter $\Lambda$. The 95\% confidence level
  lower limit on $\Lambda$ is 9.5\TeV under the assumption of
  destructive interference between the standard model and
  contact-interaction amplitudes.  For constructive interference, the
  limit is 13.1\TeV.  These limits are comparable to the most
  stringent ones reported to date.}

\hypersetup{%
pdfauthor={CMS Collaboration},%
pdftitle={Search for contact interactions in opposite-sign dimuon events in pp collisions at sqrt(s) = 7 TeV},%
pdfsubject={CMS},%
pdfkeywords={CMS, dimuons, compositeness, Contact Interactions}}

\maketitle 

\section{Introduction}

The existence of three families of quarks and leptons might be
explained if these particles are composed of more fundamental
constituents. In order to confine the constituents (often referred to
as ``preons''~\cite{PRE, Preon}) and to account for the properties of
quarks and leptons, a new strong gauge interaction, metacolor, is
introduced. Below a given interaction energy scale $\Lambda$, the
effect of the metacolor interaction is to bind the preons into
metacolor-singlet states.  For parton-parton center-of-mass energy
less than $\Lambda$, the metacolor force will manifest itself in the
form of a flavor-diagonal {\it contact
  interaction} (CI) \cite{TheoryI,TheoryII}. In the case where both
quarks and leptons share common constituents,
the Lagrangian density for a CI
leading to dimuon final states can be written as
\begin{equation}\begin{split}
\mathcal{L}_{ql} =& (g_0^2/\Lambda^2)\lbrace\eta_{LL}(\bar{\cPq}_L\gamma^\mu \cPq_L)(\bar{\mu}_L\gamma_{\mu}\mu_L)\\
                                    &+\eta_{LR}(\bar{\cPq}_L\gamma^\mu \cPq_L)(\bar{\mu}_R\gamma_{\mu}\mu_R) \\
                                     &+ \eta_{RL}(\bar{\cPqu}_R\gamma^\mu \cPqu_R)(\bar{\mu}_L\gamma_{\mu}\mu_L)\\
                                    &+\eta_{RL}(\bar{\cPqd}_R\gamma^\mu \cPqd_R)(\bar{\mu}_L\gamma_{\mu}\mu_L) \\
                                     &+ \eta_{RR}(\bar{\cPqu}_R\gamma^\mu \cPqu_R)(\bar{\mu}_R\gamma_{\mu}\mu_R)\\
                                    &+\eta_{RR}(\bar{\cPqd}_R\gamma^\mu \cPqd_R)(\bar{\mu}_R\gamma_{\mu}\mu_R) \rbrace,
\label{eq:lag}
\end{split}\end{equation}
where $\cPq_L = (\cPqu,\cPqd)_L$ is a left-handed quark doublet, $\cPqu_R$ and $\cPqd_R$
are right-handed quark singlets, and $\mu_L$ and $\mu_R$ are left- and
right-handed muons. By convention, $g_0^2/{4\pi} = 1$.  The parameter
$\Lambda$ characterizes the compositeness energy scale.  The parameters
$\eta_{ij}$ allow for differences in magnitude and phase among the
individual terms.  Lower limits on $\Lambda$ are set separately for
each term with $\eta_{ij}$ taken, by convention, to have a magnitude
of one.

\begin{figure}[b]
\begin{center}
\includegraphics[width=0.48\textwidth]{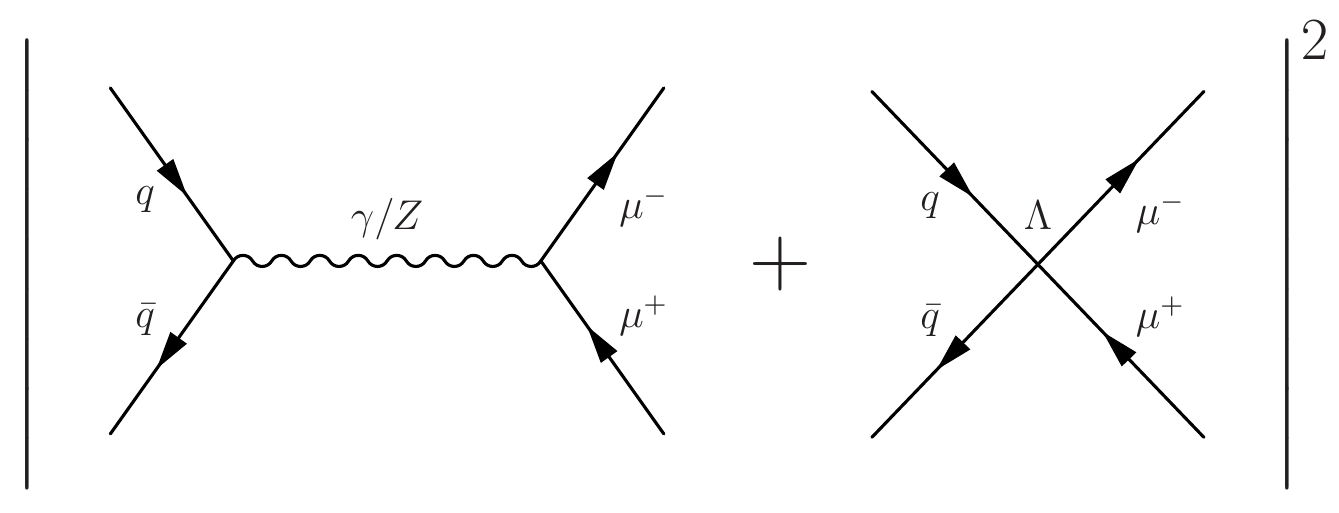}
\caption{Schematic representation of the addition of DY (left) and CI (right)
amplitudes, for common helicity states, contributing to the total
cross section for $\Pp\Pp\to X + \Pgmp\Pgmm$.}
\label{fig:interference}
\end{center}
\end{figure}

The dimuons from the subprocesses for standard model (SM) Drell--Yan
(DY) \cite{DY} production and from CI production can have the same
helicity state.  In this case, the scattering amplitudes are summed,
resulting in an interference term in the cross section for $\Pp\Pp
\to X + \Pgmp\Pgmm$, as illustrated schematically in
Fig.~\ref{fig:interference}.

The differential cross section corresponding to the combination of a
single term in Eq.~\ref{eq:lag} with DY production can be written as
\begin{equation}
\frac{\rd\sigma^\mathrm{CI/DY}}{\rd M_{\mu\mu}} =
\frac{\rd\sigma^\mathrm{DY}}{\rd M_{\mu\mu}} -
\eta_{ij}
\frac{\mathcal{I}}{\Lambda^2} + \eta_{ij}^2\frac{\mathcal{C}}{\Lambda^4},
\label{eq:cross}
\end{equation}
where $M_{\mu\mu}$ is the invariant dimuon mass, $\mathcal{I}$ is due to
interference, and ${\mathcal C}$ is purely due to the CI.  Note that
$\eta_{ij}=+1$ corresponds to destructive interference and
$\eta_{ij}=-1$ to constructive interference. The processes
contributing to the cross section in Eq.~(\ref{eq:cross}) are denoted
collectively by ``CI/DY''.  The difference $\rd\sigma^{\rm
  CI/DY}/\rd M_{\mu\mu} - \rd\sigma^\mathrm{DY}/\rd M_{\mu\mu}$ is the signal we
are searching for in this paper.

The contact interaction model used for this analysis is the left-left
isoscalar model (LLIM) \cite{TheoryII}, which corresponds to a
left-handed current interaction described by the first term of
$\mathcal{L}_{\cPq l}$ in Eq.~(1). The LLIM is the conventional benchmark
for CI in the dilepton channel. For this analysis, all initial-state
quarks are assumed to be composite.

Previous searches for CI in the dijet and dilepton channels have all
resulted in limits on the compositeness scale $\Lambda$. Searches have
been reported from experiments at LEP
~\cite{ALEPH,DELPHI,L3,OPAL1,OPAL2}, HERA~\cite{H1,ZEUS}, the
Tevatron~\cite{CDF1,CDF2,CDF3,CDF4,D01,D02}, and recently from the
ATLAS~\cite{ATLAS1,ATLAS2,ATLAS3,ATLAS4} and
CMS~\cite{CMS1,CMS_dijet,CMSBOOSTEDZ} experiments at the LHC. The best
limits in the LLIM dimuon channel are $\Lambda > 9.6$\TeV for
destructive interference and $\Lambda > 12.9$\TeV for constructive
interference, at the 95\% confidence level (CL)~\cite{ATLAS4}.

In this paper, we report a search for CI in the dilepton channel
produced in $\Pp\Pp$ collisions at $\sqrt{s} = 7$\TeV using the Compact
Muon Solenoid (CMS) detector at the Large Hadron Collider (LHC). The
data sample corresponds to an integrated luminosity of 5.3\fbinv.

\section{Predictions of the left-left isoscalar model}

\begin{figure}[b]
\begin{center}
         \includegraphics[width=0.45\textwidth]{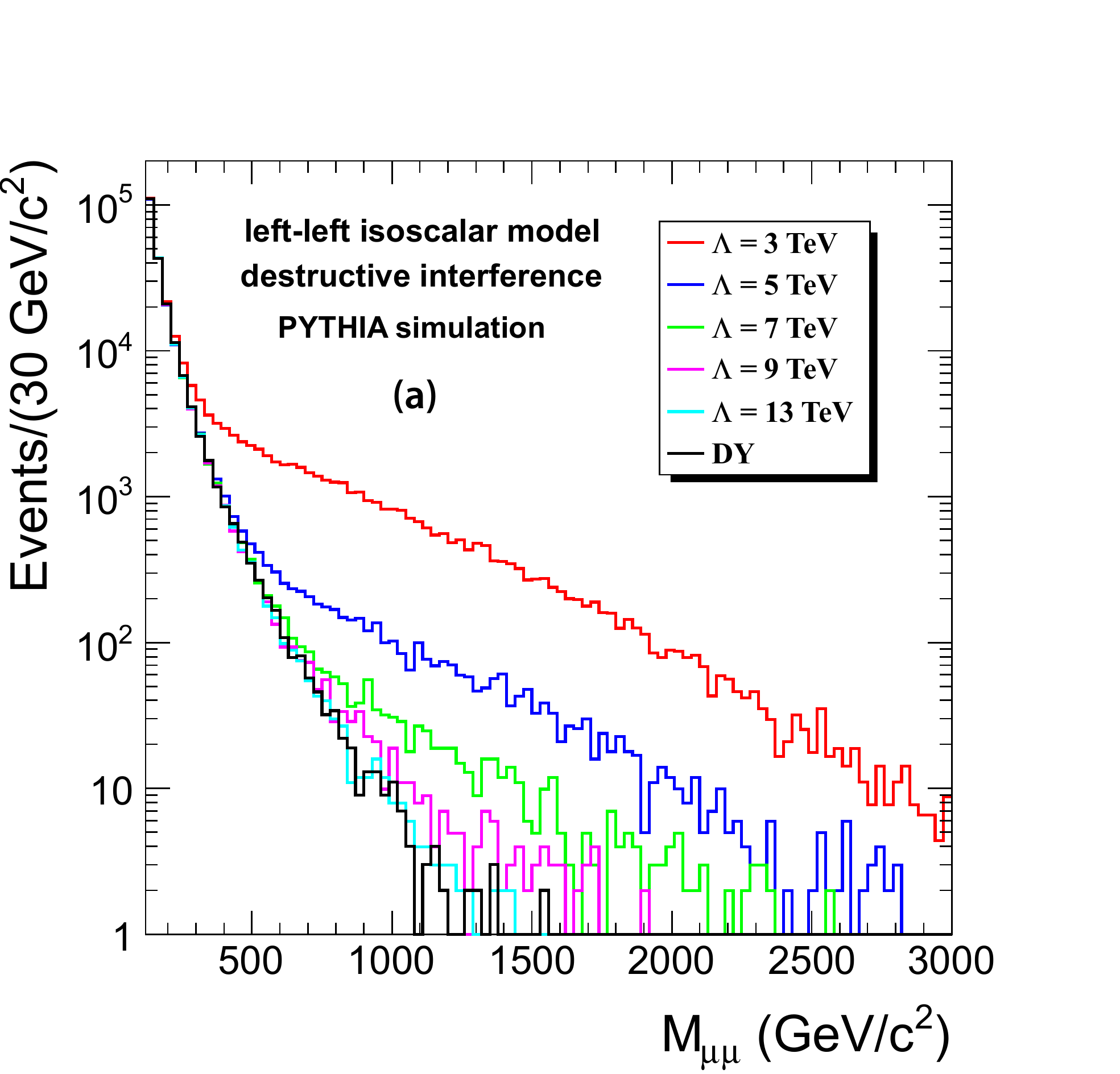}
         \includegraphics[width=0.45\textwidth]{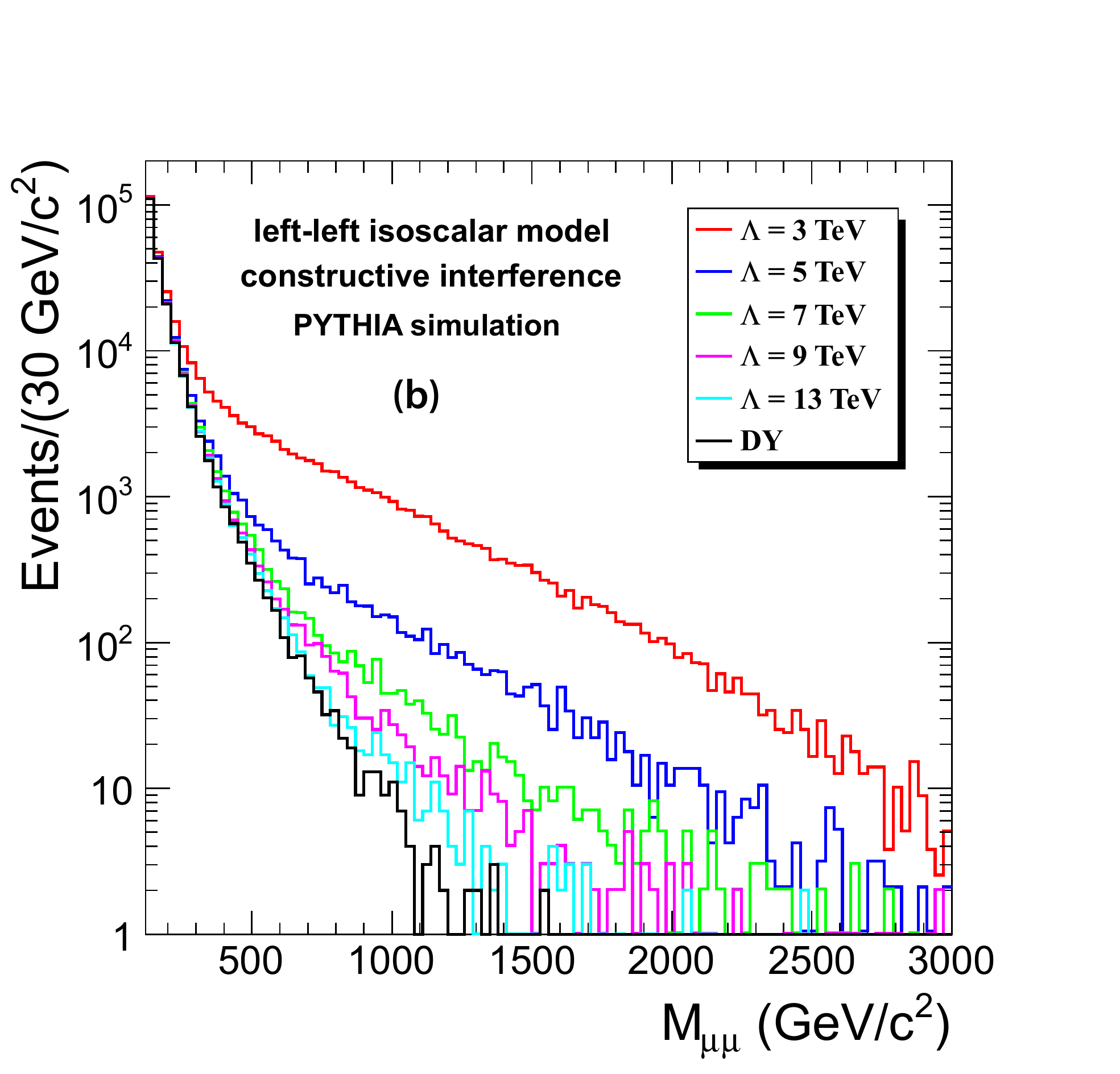}
  \caption
  {Simulated dimuon mass spectra using the left-left isoscalar model
    for different values of $\Lambda$ for (a) destructive interference
    and (b) constructive interference.  The events are generated using
    the {\PYTHIA} Monte Carlo program with kinematic selection
    requirements that approximate the acceptance of the detector.  As
    $\Lambda$ increases, the effects of the CI are reduced, and the
    event yield approaches that for DY production.  The model
    predictions are shown over the full mass range, although the model
    is not valid as $M_{\mu\mu}c^2$ approaches $\Lambda$. The
    integrated luminosity corresponds to 63\fbinv. \label{fig:dimuon}}

\end{center}
\end{figure}

The basic features of the LLIM dimuon mass spectra are demonstrated
with a generator-level simulation using {\PYTHIA} \cite{{PYTHIA}},
with appropriate kinematic selection criteria that approximate the
acceptance of the detector.
Figures~\ref{fig:dimuon}(a) and \ref{fig:dimuon}(b) show the LLIM
dimuon mass spectra for different values of $\Lambda$ for destructive
and constructive interference, respectively. The curves illustrate
that with increasing mass the CI leads to a less steeply falling yield
relative to DY production, with the effect steadily increasing with
decreasing $\Lambda$. For a given value of $\Lambda$, the event yield
is seen to be larger for constructive interference compared to the
destructive case, with the relative difference increasing with
$\Lambda$.

For the results presented in this paper, the analysis
is limited to a dimuon mass range from 200 to 2000\GeVcc. The lower
mass is sufficiently above the \cPZ-peak so that a deviation from DY
production would be observable.  The highest dimuon mass observed is
between 1300 and 1400\GeVcc and, for the values of $\Lambda$ where
the limits are set, less than one event is expected for dimuon masses
above 2000\GeVcc. In order to limit the mass range in which the
detector acceptance has to be evaluated, we therefore choose an upper
mass cutoff of 2000\GeVcc.  To optimize the limit on $\Lambda$, a
minimum mass $M_{\mu\mu}^{\min}$ is varied between the lower and
upper mass values, as described in Section \ref{Limits}.

\section{CMS detector}

The central feature of the CMS apparatus is a
superconducting solenoid of 6\unit{m} internal diameter, providing a
magnetic field of 3.8\unit{T}. Within the field volume are a silicon
pixel and strip tracker, a lead tungstate crystal electromagnetic
calorimeter, and a brass/scintillator hadron calorimeter. Muons are
measured in gas-ionization detectors embedded in the steel flux-return
yoke. Extensive forward calorimetry complements the coverage provided
by the barrel and endcap detectors.  A detailed description of the CMS
detector can be found in Ref.~\cite{CMS2}.

The tracker and muon detector are important sub-systems for this
measurement.  The tracker measures charged particle trajectories
within the range $|\eta| <$ 2.5, where pseudorapidity
$\eta = -\ln[\tan(\theta/2)]$,
and polar angle $\theta$ is
measured from the beam axis.  The tracker provides a transverse
momentum ($\pt$) resolution of about 1\% at a few tens of\GeVc
to 10\% at several hundred\GeVc~\cite{PTRES}, where $\pt$ is
the component of momentum in the plane perpendicular to the beam axis.
Tracker elements include about 1400 silicon pixel modules, located
close to the beamline, and about 15\,000 silicon microstrip modules,
which surround the pixel system.  Tracker detectors are arranged in
both barrel and endcap geometries.  The muon detector comprises a
combination of drift tubes and resistive plate chambers in the barrel
region and a combination of cathode strip chambers and resistive plate
chambers in the endcap regions. Muons can be reconstructed in the
range $|\eta| < 2.4$.

For the trigger path used in this analysis, the first level (L1)
selects events with a muon candidate based on a sub-set of information
from the muon detector.  The trigger muon is required to have $|\eta|
< 2.1$ and $\pt$ above a threshold that was raised to 40\GeVc by the end of the data-taking period.  This cut has little
effect on the acceptance for muon pairs with masses above 200\GeVcc. The small
effect is included in the simulation.  The high level trigger (HLT)
refines the L1 selection using the full information from both the
tracker and muon systems.

\section{Event selection criteria}

This analysis uses the same event selection as the search for new
heavy resonances in the dimuon channel, discussed in
Ref.~\cite{ZPRIME12}.
Each muon track is required to have a signal
(``hit'') in at least one pixel layer, hits in at least nine strip
layers, and hits in at least two muon detector stations.  Both muons
are required to have $\pt >  45\GeVc$. To reduce the cosmic ray
background, the transverse impact parameter of the muon with respect
to the beamspot is required to be less than 0.2\cm.  In order to
suppress muons coming from hadronic decays, a tracker-based isolation
requirement is imposed such that the sum of $\pt$ of all tracks,
excluding the muon and within a cone surrounding the muon, is less
than 10\% of the $\pt$ of the muon.  The cone is defined by the
condition $\Delta R$ = $\sqrt{(\Delta \eta)^2 + (\Delta \phi)^2} = 0.3$, where
$\phi$ is the azimuthal angle of a track, and the differences
$\Delta \eta$ and $\Delta \phi$
are determined with respect to the muon's direction.

The two muons are required to have opposite charge and to be
consistent with originating from a common vertex.  To suppress cosmic
ray muons that are in time with the collision event, the angle between
the two muons must be smaller than $\pi - 0.02$ radians.  At least one
of the reconstructed muons must be matched (within $\Delta R<$ 0.2 and
$\Delta \pt/\pt<$ 1) to the HLT muon candidate.

If an event has more than two muons passing the above requirements,
then the two highest-$\pt$ muons are selected, and the event is
retained only if these muons are oppositely charged. Only three such
events are observed with selected dimuon mass above 200\GeVcc, and in all
three cases, the dimuon mass is less than 300\GeVcc. Thus, events
with multiple dimuon candidates play essentially no role in the
analysis.

\section{Simulation of SM and CI dimuon production}

This section describes the method used to simulate the mass
distribution from the CI/DY process of Eq.~(\ref{eq:cross}), including
the leading-order (LO) contributions from DY and CI amplitudes, their
interference, the effects of next-to-leading-order (NLO) QCD and QED
corrections, and the response of the detector.  The predicted number
of CI/DY events is the product of the generated number of CI/DY
events, a QCD K-factor, a QED K-factor, and a factor denoted as
``acceptance times migration'' ($A \times M$). The factor $A \times M$
is determined from the detector simulation of DY events, as explained
below in Section \ref{accept_ci}.  The simulation of background due to
non-DY SM processes is also described.

\subsection{Event samples with detector simulation}

A summary of the event samples used for simulation of the detector
response to various physics processes is presented in Table
\ref{tab:dset_det}.  The event generators used are {\PYTHIA}, with the
CTEQ6.6M implementation \cite{CTEQ66M} of parton distribution
functions (PDF), {\POWHEG}~\cite{PHEG1,PHEG2,PHEG3}, and
\MADGRAPH{5}~\cite{MADGRAPH}. The detector simulation is based on
\GEANTfour \cite{GEANT}.

\begin{table*} [tb]\centering
    \topcaption{Description of event samples with detector simulation.
      The cross section $\sigma$ and integrated luminosity $L$ are given for
      each sample generated.
    }

  \begin{footnotesize}
  \scotchrule[ccD{,}{\times}{4.4}D{,}{\times}{4.4}D{,}{\times}{4.4}c]
    Process  & Generator                                             & \multicolumn{1}{c}{\# of Events}       & \multicolumn{1}{c}{$\sigma$(pb)} & \multicolumn{1}{c}{$L$(\pbinv)} & Order \\
    \hline
    \rule{0pt}{2.5ex}$\cPZ/\gamma^{*}\to\mu\mu$, $M_{\mu\mu} \geq 120$\GeVcc & {\PYTHIA} & 5.45 , 10^{4}  &  7.90  , 10^{0}	   & 6.91 , 10^{3} & LO     \\
    $\cPZ/\gamma^{*}\to\mu\mu$, $M_{\mu\mu} \geq 200$\GeVcc & {\PYTHIA} & 5.50 , 10^{4}  &  9.70  , 10^{-1}    & 5.67 , 10^{4} & LO     \\
    $\cPZ/\gamma^{*}\to\mu\mu$, $M_{\mu\mu} \geq 500$\GeVcc & {\PYTHIA} & 5.50 , 10^{4}  &  2.70  , 10^{-2}    & 2.04 , 10^{6} & LO     \\
    $\cPZ/\gamma^{*}\to\mu\mu$, $M_{\mu\mu} \geq 800$\GeVcc & {\PYTHIA} & 5.50 , 10^{4}  &  3.10  , 10^{-3}    & 1.77 , 10^{7} & LO     \\
    $\cPZ/\gamma^{*}\to\mu\mu$, $M_{\mu\mu} \geq 1000$\GeVcc& {\PYTHIA} & 5.50 , 10^{4}  &  9.70  , 10^{-4}    & 5.67 , 10^{7} & LO     \\
    $\cPZ/\gamma^{*}\to\tau\tau$                            & {\PYTHIA} & 2.03 , 10^{6}  &  1.30  , 10^{3}     & 1.56 , 10^{3} & LO     \\
    $\ttbar$ & {\MADGRAPH}                                              & 2.40 , 10^{6}  &  1.57  , 10^{2}     & 1.54 , 10^{5} & NLO    \\
    $\cPqt\PW$ & {\POWHEG}                                              & 7.95 , 10^{5}  &  7.90  , 10^{0}	   & 1.01 , 10^{5} & NLO    \\
    $\cPaqt\PW$ & {\POWHEG}                                             & 8.02 , 10^{5}  &  7.90  , 10^{0}	   & 1.02 , 10^{5} & NLO    \\
    $\PW\PW$ & {\PYTHIA}                                                & 4.23 , 10^{6}  &  4.30  , 10^{1}	   & 9.83 , 10^{4} & LO     \\
    $\PW\cPZ$ & {\PYTHIA}                                               & 4.27 , 10^{6}  &  1.80  , 10^{1}	   & 2.37 , 10^{5} & LO     \\
    $\cPZ\cPZ$ & {\PYTHIA}                                              & 4.19 , 10^{6}  &  5.90  , 10^{0}	   & 7.10 , 10^{5} & LO     \\
    $\PW$ + jets & {\MADGRAPH}                                          & 2.43 , 10^{7}  &  3.10  , 10^{4}     & 7.82 , 10^{2} & NLO    \\
    multi-jet, $\mu$ ($\pt > 15$\GeVc) & {\PYTHIA}                      & 1.08 , 10^{6}  &  8.47  , 10^{4}     & 1.28 , 10^{2} & LO     \\
  \donescotchrule
    \end{footnotesize}
  \label{tab:dset_det}
\end{table*}

\subsection{Detector acceptance times mass migration}
\label{accept_ci}

\begin{figure}[bt]
\begin{center}
         \includegraphics[width=0.5\textwidth]{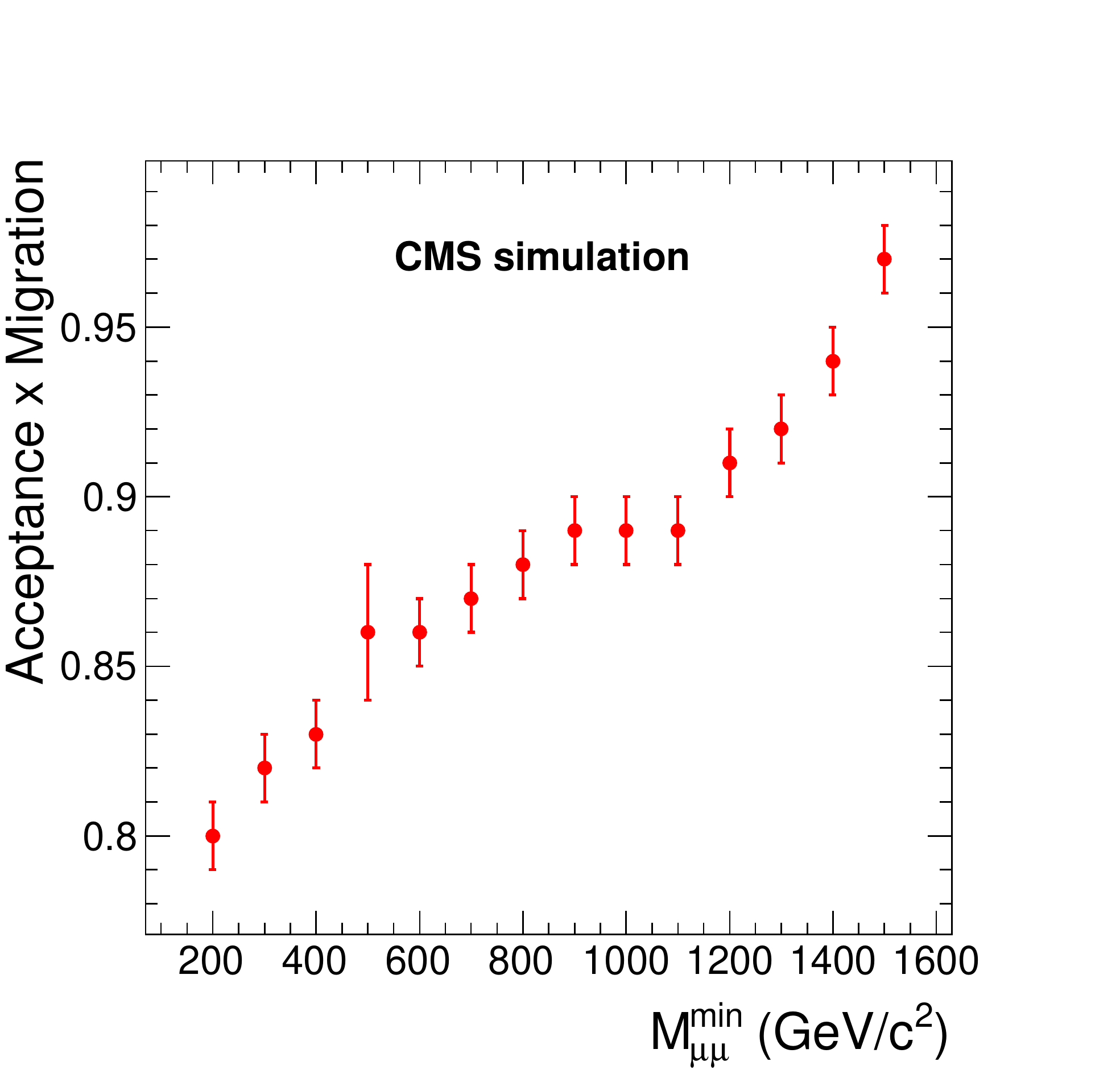}
  \caption{Acceptance times migration, $A \times M$, versus
    $M_{\mu\mu}^{\min}$.  Corresponding values and uncertainties
    are given in Table \ref{tab:factors}.  The error bars indicate
    statistical uncertainties based on simulation of the DY process.
    The systematic uncertainty is 3\% as explained in the text.  The
    increase of $A \times M$ at lower mass is due to the increase in
    acceptance, while at higher mass, it is dominated by the growth in
    mass resolution.  Since the cross section falls steeply with mass,
    events tend to migrate from lower to higher mass over a range
    determined by the mass resolution.}

\label{fig:accept_comparison}
\end{center}
\end{figure}

To simplify the analysis, we use the detector simulation for DY events
to determine the detector response for CI/DY events, which have a
behavior similar to that for DY events for the large values of
$\Lambda$ of interest in this analysis.  For a given value of
$M_{\mu\mu}^{\min}$, the product of acceptance times migration ($A
\times M$) is given by the ratio of the number of DY events
reconstructed with mass above $M_{\mu\mu}^{\min}$ to the number of
DY events generated with mass above $M_{\mu\mu}^{\min}$. Some of
the reconstructed events have been generated with mass below
$M_{\mu\mu}^{\min}$ because of the smearing due to the mass
reconstruction, which has a resolution of 6.5\% at masses around 1000\GeVcc, rising to 12\% at 2000\GeVcc.  The dependence of $A \times M$
on $M_{\mu\mu}^{\min}$ is plotted in
Fig.~\ref{fig:accept_comparison} and values are given in Table
\ref{tab:factors}.  The increase of $A \times M$ at lower mass is due
to the increase in acceptance, while at higher mass, it is dominated
by the growth in mass resolution.  Since the cross section falls
steeply with mass, events tend to migrate from lower to higher mass
over a range determined by the mass resolution.

\begin{table} [tb]\centering
  \topcaption{Multiplicative factors used in the prediction of
    the expected number of events from the CI/DY process. The uncertainties
    shown are statistical. The systematic uncertainty is 3\% for $A \times M$
    and 3\% for the QCD K-factor, as explained in the text.
    The uncertainty in the QED K-factor is dominated by the systematic
    uncertainty that is assigned as the size
    of the correction, $\abs{\text{(QED K-factor)} -1}$, to allow for systematic
    uncertainty in the generator.}
    \begin{footnotesize}
      \scotchrule[cD{,}{\,\pm\,}{4.4}D{,}{\,\pm\,}{4.4}c]
        $M_{\mu\mu}^{\min}$(\GeVccns{})   & \multicolumn{1}{c}{$A \times M$}  & \multicolumn{1}{c}{QCD K-factor}
        & QED K-factor \\
        \hline
        200   & 0.80 , 0.01   &1.303 ,  0.005   & 1.01  \\
        300   & 0.82 , 0.01   &1.308 ,  0.005   & 0.99  \\
        400   & 0.83 , 0.01   &1.299 ,  0.005   & 0.97  \\
        500   & 0.86 , 0.02   &1.305 ,  0.005   & 0.95  \\
        600   & 0.86 , 0.01   &1.299 ,  0.005   & 0.94  \\
        700   & 0.87 , 0.01   &1.298 ,  0.005   & 0.92  \\
        800   & 0.88 , 0.01   &1.288 ,  0.005   & 0.91  \\
        900   & 0.89 , 0.01   &1.280 ,  0.004   & 0.90  \\
        1000   & 0.89 , 0.01   &1.278 ,  0.004   & 0.89  \\
        1100   & 0.89 , 0.01   &1.275 ,  0.004   & 0.88  \\
        1200   & 0.91 , 0.01   &1.268 ,  0.004   & 0.88  \\
        1300   & 0.92 , 0.01   &1.262 ,  0.004   & 0.87  \\
        1400   & 0.94 , 0.01   &1.260 ,  0.004   & 0.87  \\
        1500   & 0.97 , 0.01   &1.261 ,  0.004   & 0.86  \\
      \donescotchrule
    \end{footnotesize}

    \label{tab:factors}
  \end{table}

To validate that the $A \times M$ factor based on DY production is
  applicable to CI/DY production, we compare event yields predicted
  using the $A \times M$ factor with those predicted using a simulation
  of CI/DY production.  The study is performed for the cases of
  constructive interference with $\Lambda = 5$ and 10\TeV, which
  represent a wide range of possible CI/DY cross sections.  The results
  differ by at most 3\%, consistent with the statistical precision of
  the study. The systematic uncertainty in
  $A \times M$ is conservatively assigned this value.

\subsubsection{Event pileup}

During the course of the 2011 data taking period, the luminosity
increased with time, resulting in an increasing ``event pileup'', the
occurrence of multiple $\Pp\Pp$ interactions recorded by the detector as a
single event.  The dependence of reconstruction efficiency on event
pileup is studied by weighting simulated events so that the
distribution of the number of reconstructed primary vertices per event
matches that in data. The reconstruction efficiency is found to be
insensitive to the variations in event pileup encountered during the
data taking period.

\subsection{Higher-order strong and electromagnetic corrections}
\label{kfactor}

Since we use the leading-order generator {\PYTHIA} to simulate the
CI/DY production, we must determine a QCD K-factor which takes into
account higher-order initial-state diagrams.  Under the assumption
that the QCD K-factor is the same for DY and CI/DY events, we
determine the QCD K-factor as the ratio of DY events generated using
the next-to-leading-order generator {\sc MC@NLO}~\cite{MCNLO} to those
generated using {\PYTHIA}.  The {\MCATNLO} generator is used with
the same PDF set as used with {\PYTHIA}.  The resulting QCD
K-factor as a function of $M_{\mu\mu}^{\min}$ is given in Table
\ref{tab:factors}.  The large sizes of the simulated event samples
result in statistical uncertainties less than 0.5\%. The systematic
uncertainty is assigned the value 3\%, the size of the correction
\cite{NNLO} between next-to-next-to-leading order (NNLO) and NLO DY
cross sections.  For SM processes other than DY production, the QCD
K-factor is found, independent of dimuon mass, from the ratio of the
cross section determined using {\MCATNLO} to the cross section
determined from {\PYTHIA}.

The effect of higher-order electromagnetic processes on CI/DY
production is quantified by a mass-dependent QED K-factor determined
using the HORACE generator \cite{EWK}. The values of the QED K-factor,
as a function of $M_{\mu\mu}^{\min}$, are given in Table
\ref{tab:factors}. The systematic uncertainty is assigned as the size
of the correction, $\abs{\text{(QED K-factor)} -1}$, since the effect of
higher-order QED corrections on the new physics of CI is unknown.

\subsection{Non-DY SM backgrounds}

Using the samples of simulated events listed in Table
\ref{tab:dset_det}, event yields are predicted for various non-DY SM
background processes, as shown in Table \ref{tab:bg}.  The yields are
given as a function of $M_{\mu\mu}^{\min}$ and they are scaled to
the integrated luminosity of the data, $5.28\pm0.12\fbinv$
\cite{LUMI}.  For comparison, the expected yields are also shown for
DY events.  The relevant backgrounds, in decreasing order of
importance, are $\ttbar$, diboson (\PW\PW/\PW\cPZ/\cPZ\cPZ), \PW\ (including
\PW+jets and \cPqt\PW), and $\cPZ\to\Pgt\Pgt$ production.  The background
from multi-jet events is studied using both the simulation sample
listed in Table \ref{tab:dset_det} and control samples from data, as
reported in Ref.~\cite{ZPRIME12}.  The results of either method
indicate that no multi-jet background events are expected for
$M_{\mu\mu}^{\min} > 200\GeVcc$.  For $M_{\mu\mu}^{\min} >
1000$\GeVcc the fractional statistical uncertainty in the non-DY
background is large, but the absolute yield is much smaller than that
for DY background.

\begin{table*} [bt]\centering
  \topcaption{Expected event yields for DY and non-DY
    SM backgrounds. The uncertainties shown are statistical.
    A systematic uncertainty of 2.2\% arises from the determination of
    integrated luminosity
    \cite{LUMI}.}
  \begin{footnotesize}
  \scotchrule[cD{,}{\,\pm\,}{5.4}D{,}{\,\pm\,}{4.4}D{,}{\,\pm\,}{4.4}D{,}{\,\pm\,}{5.4}D{,}{\,\pm\,}{4.4}D{,}{\,\pm\,}{5.4}]

\multicolumn{1}{c}{$M_{\mu\mu}^{\min}$} & \multicolumn{1}{c}{DY} & \multicolumn{1}{c}{\ttbar} & \multicolumn{1}{c}{Diboson} & \multicolumn{1}{c}{\PW+Jets \& \cPqt\PW} &
\multicolumn{1}{c}{\cPZ$\rightarrow\Pgt\Pgt$}
    & \multicolumn{1}{c}{Sum Non-DY} \\
\multicolumn{1}{c}{(\GeVccns{})}  & &&&&&\\

\hline
  200 &  3630   ,  18    & 454     ,  3    & 123.0   ,  2    & 47.90  ,  1.35 & 6.96  ,  4.14 & 632.3   ,  5.9  \\
  300 &  870.6  ,   8.8  & 104     ,  2    &  38.6   ,  1.2  & 12.82  ,  0.70 & \multicolumn{1}{c}{0}         & 155.9   ,  2.1  \\
  400 &  301.6  ,   5.1  &  26.0   ,  0.8  &  12.7   ,  0.7  &  3.32  ,  0.35 & \multicolumn{1}{c}{0}         &  42.0   ,  1.1  \\
  500 &  123.8  ,   3.3  &   8.19  ,  0.46 &   5.07  ,  0.41 &  1.02  ,  0.20 & \multicolumn{1}{c}{0}         &  14.3   ,  0.6  \\
  600 &  55.31  ,   0.19 &   2.92  ,  0.27 &   2.42  ,  0.28 &  0.29  ,  0.11 & \multicolumn{1}{c}{0}         &   5.63  ,  0.41 \\
  700 &  27.35  ,   0.13 &   1.12  ,  0.17 &   0.86  ,  0.16 &  0.07  ,  0.05 & \multicolumn{1}{c}{0}         &   2.06  ,  0.24 \\
  800 &  14.23  ,   0.10 &   0.34  ,  0.09 &   0.51  ,  0.12 &  0.07  ,  0.05 & \multicolumn{1}{c}{0}         &   0.92  ,  0.16 \\
  900 &   7.72   ,   0.07 &   0.05  ,  0.03 &   0.25  ,  0.08 &  0.07  ,  0.05 & \multicolumn{1}{c}{0}        	&   0.36  ,  0.10 \\
 1000 &   4.32  ,   0.05 &   0.05  ,  0.03 &   0.10  ,  0.05 &  0.07  ,  0.05 & \multicolumn{1}{c}{0}         &   0.21  ,  0.08 \\
 1100 &   2.46  ,   0.04 &   0.05  ,  0.03 &   0.09  ,  0.05 &  0.07  ,  0.05 & \multicolumn{1}{c}{0}         &   0.20  ,  0.08 \\
 1200 &   1.48  ,   0.03 &   \multicolumn{1}{c}{0} 	       &   0.01  ,  0.01 &  0.07  ,  0.05 & \multicolumn{1}{c}{0}         &   0.08  ,  0.05 \\
 1300 &   0.91  ,   0.02 &   \multicolumn{1}{c}{0}        &   0.01  ,  0.01 &  0.07  ,  0.05 & \multicolumn{1}{c}{0}         &   0.08  ,  0.05 \\
 1400 &   0.56  ,   0.02 &   \multicolumn{1}{c}{0}        &   0.01  ,  0.01 &  0.07  ,  0.05 & \multicolumn{1}{c}{0}         &   0.08  ,  0.05 \\
 1500 &   0.33  ,   0.02 &   \multicolumn{1}{c}{0}        &   \multicolumn{1}{c}{0}            &  0.07  ,  0.05 & \multicolumn{1}{c}{0}         &   0.07  ,  0.05 \\
  \donescotchrule
  \end{footnotesize}   \label{tab:bg}
\end{table*}

\subsection{Predicted event yields}

\label{sect:pred_evt_yields}

Using the methods described above,
the sum of the event yields for the
CI/DY process and the non-DY SM backgrounds,
for the integrated luminosity of the data sample, are predicted as a
function of $M_{\mu\mu}^{\min}$ and $\Lambda$.
The predicted event yields for
destructive and constructive interference are given in Tables
\ref{tab:reco_dest} and \ref{tab:reco_cons}.

For destructive interference, there is a region of the
$M_{\mu\mu}^{\min}$-$\Lambda$ parameter space where the predicted
number of events
is less than for SM production.
This ``reduced-yield'' region is indicated in Table~\ref{tab:reco_dest}.  The region of parameter space, $M_{\mu\mu}^{\min} > 600$\GeVcc and $\Lambda \leq 12$\TeV, where our expected
limit is most stringent (see Fig.  \ref{fig:cl}(a)), lies outside
the reduced-yield region.  For constructive interference, the
predicted number of events is always larger than
for SM production.

\begin{table*}[tb]\centering
  \topcaption{Observed and expected number of events for illustrative
    values of $ M_{\mu\mu}^{\min}$.  The expected yields are shown
    for SM production and for the sum of CI/DY production (for
    destructive interference and for a given $\Lambda$) and non-DY SM
    backgrounds.  For each column of $M_{\mu\mu}^{\min}$, the
    expected yield for CI/DY + non-DY SM production that is just \textit{above} that expected for SM production is highlighted in red.
    Entries above the highlighted ones correspond to values of
    $\Lambda$ for which the expected yield is \textit{less} than that for
    SM production, because of the destructive interference term in the
    cross section.
    As discussed in Section \ref{sect:results}, the best expected limit is
    obtained for $M^{\min}_{\mu\mu}$ = 1100\GeVcc. For this choice,
    the expected event yield is highlighted in gray that corresponds
    to the value of $\Lambda$ closest to the
    observed 95\% CL lower limit on $\Lambda$ of 9.5\TeV (9.7\TeV
    expected).}
  \begin{scriptsize}
  \scotchrule[cD{.}{.}{4.1}D{.}{.}{4.1}D{.}{.}{4.1}D{.}{.}{4.1}D{.}{.}{4.1}D{.}{.}{4.1}D{.}{.}{4.1}D{.}{.}{4.1}D{.}{.}{4.1}D{.}{.}{4.1}D{.}{.}{4.1}]
$M_{\mu\mu}^{\min}$ &\multicolumn{1}{c}{}&\multicolumn{1}{c}{}&\multicolumn{1}{c}{}&\multicolumn{1}{c}{}&\multicolumn{1}{c}{}&\multicolumn{1}{c}{}&\multicolumn{1}{c}{}&\multicolumn{1}{c}{}&\multicolumn{1}{c}{}&\multicolumn{1}{c}{}&\multicolumn{1}{c}{} \\
(\GeVccns{})       &
\multicolumn{1}{c}{500}        &    \multicolumn{1}{c}{600}        &    \multicolumn{1}{c}{700}        &    \multicolumn{1}{c}{800}        &    \multicolumn{1}{c}{900}        &    \multicolumn{1}{c}{1000}
      &    \multicolumn{1}{c}{1100}       & \multicolumn{1}{c}{1200}  & \multicolumn{1}{c}{1300} & \multicolumn{1}{c}{1400} & \multicolumn{1}{c}{1500} \\
\hline
\multicolumn{1}{c}{Source}  &  \multicolumn{11}{c} {Number of Events} \\
\hline

data &    141   &    57   &    28    &    14    &   13     &    8     &    3     &    2     &    1     &    0     &    0   \\

SM MC  &       138.1    &      60.9     &      29.4    &       15.2    &       8.1     &       4.5     &       2.7     &       1.6     &       1.0     &       0.6     &       0.4   \\

$\Lambda$ (\TeVns{}) MC        & &               &             &               &
      &               &               &               &               &
 &               \\

18	&	134.2	&	58.0	   &	27.9	    &	14.3	    &	7.7	    &	4.3	    &	2.6		&	1.5	   &	1.0	    &	0.6		&	0.4		\\
17	&	134.5	&	57.9	   &	27.7	    &	14.4	    &	7.8	    &	4.4	    &	2.6		& \cellcolor{Salmon}\textbf{1}.\textbf{6}	   &	1.0	    &	0.6	&	0.4		\\
16	&	134.9	&	58.0	   &	27.8	    &	14.5	    &	7.8	    &	4.5	    &	\cellcolor{Salmon}\textbf{2}.\textbf{7}		&	1.6	   &	1.0	    &	0.7		&	0.5		\\
15	&	135.6	&	58.3	   &	28.1	    &	14.7	    &	8.0	    &	\cellcolor{Salmon}\textbf{4}.\textbf{7}	    &	2.9		&	1.7	   &	1.1	    &	0.8		&	0.5		\\
14	&	133.7	&	58.3	   &	28.3	    &	15.0	    &	\cellcolor{Salmon}\textbf{8}.\textbf{4}	    &	5.0	    &	3.1		&	1.9	   &	1.3	    &	0.9		&	0.6		\\
13	&	134.1	&	59.3	   &	29.1	    &	\cellcolor{Salmon}\textbf{15}.\textbf{7}	    &	8.9	    &	5.4	    &	3.5		&	2.2	   &	1.5	    &	1.0		&	0.7		\\
12	&	138.6	&	60.1	   &	\cellcolor{Salmon}\textbf{30}.\textbf{2}	    &	16.7	    &	9.8	    &	6.1	    &	4.1		&	2.7	   &	1.9	    &	1.3		&	0.9		\\
11	&	135.7	&	\cellcolor{Salmon}\textbf{62}.\textbf{5}	   &	32.1	    &	18.4	    &	11.2	    &	7.3	    &	5.0		&	3.5	   &	2.4	    &	1.7		&	1.2		\\
10	&	\cellcolor{Salmon}\textbf{141}.\textbf{1}	&	66.7	   &	35.7	    &	21.2	    &	13.6	    &	9.2	
& \cellcolor{lightgray}6.6		&	4.6	   &	3.3	    &	2.4		&	1.7		\\
9	&	148.5	&	73.8	   &	42.4	    &	27.1	    &	18.3	    &	13.1	    &	9.5		&	6.9	   &	5.0	    &	3.7		&	2.6		\\
8	&	164.7	&	88.1	   &	54.4	    &	36.8	    &	26.2	    &	19.3	    & 14.3 &	10.6	   &	7.8	    &	5.8		&	4.1 \\		
7	&	198.1	&	117.5  &	79.4	    &	57.6	    &	43.3	    &	31.6	    &	24.0		&	17.5	   &	13.1	    &	9.0		&	6.2		\\
6	&	278.1	&	182.3  &	131.7   &	100.1   &	76.7	    &	57.9	    &	45.1		&	33.0	   &	22.6	    &	16.9		&	11.3		\\
5	&	469.2	&	338.7  &	261.7   &	204.4   &	158.6   &	123.2   &	96.7		&	74.6	   &	56.8	    &	41.5		&	29.2		\\
4	&	1025	&	784.1  &	620.1   &	494.2   &	384.3   &	302.6   &	232.8	&	174.6  &	127.2   &	94.5		&	68.5		\\
3	&	3199	& 2517  &  2012    &	1599	    &	1242	    &	~975.7  &	744.7	&	575.4  &	437.7   &	320.1	&	231.4	\\

  \donescotchrule
  \end{scriptsize}

  \label{tab:reco_dest}
\end{table*}

\begin{table*}[tb]\centering
  \topcaption{Observed and expected number of events as in
    Table \ref{tab:reco_dest}.  Here CI/DY predictions are for
    constructive interference.  Shown with gray highlighting is the
    expected event yield corresponding to the value of $\Lambda$
    closest to the observed 95\% CL lower limit on $\Lambda$ of 13.1\TeV (12.9\TeV expected) for $M^{\min}_{\mu\mu}$ selected to be
    800\GeVcc.}
  \begin{scriptsize}
  \scotchrule[cD{.}{.}{4.1}D{.}{.}{4.1}D{.}{.}{4.1}D{.}{.}{4.1}D{.}{.}{4.1}D{.}{.}{4.1}D{.}{.}{4.1}D{.}{.}{4.1}D{.}{.}{4.1}D{.}{.}{4.1}D{.}{.}{4.1}]
$M_{\mu\mu}^{\min}$ &&&&&&&&&&& \\

(\GeVccns{})       &       \multicolumn{1}{c}{400}     &
\multicolumn{1}{c}{500}        &    \multicolumn{1}{c}{600}        &    \multicolumn{1}{c}{700}        &    \multicolumn{1}{c}{800}        &    \multicolumn{1}{c}{900}        &    \multicolumn{1}{c}{1000}
      &    \multicolumn{1}{c}{1100}       & \multicolumn{1}{c}{1200}  & \multicolumn{1}{c}{1300} & \multicolumn{1}{c}{1400} \\
\hline
Source  &  \multicolumn{11}{c} {Number of Events} \\
\hline

data &    338   &    141   &    57    &    28    &    14    &   13     &    8     &    3     &    2     &    1     &    0     \\

SM MC  &       343.6   &       138.1    &      60.9     &      29.4    &       15.2    &       8.1     &       4.5     &       2.7     &       1.6     &       1.0     &       0.6    \\

$\Lambda$ (\TeV) MC        & &               &             &               &
      &               &               &               &               &
 &               \\

18	&	359.2	&	147.7	&	67.8	   &	34.1	   &	18.7	   &	10.6	   &	6.4	   &	4.0	   &	2.5	   &	1.6	   &	1.1		\\
17	&	358.9	&	149.3	&	69.1	   &	35.1	   &	19.3	   &	11.1	   &	6.7	   &	4.3	   &	2.7	   &	1.8	   &	1.2		\\
16	&	365.2	&	153.7	&	70.3	   &	36.1	   &	20.2	   &	11.7	   &	7.2	   &	4.6	   &	3.0	   &	2.0	   &	1.3		\\
15	&	365.6	&	156.3	&	71.9	   &	37.2	   &	20.9	   &	12.3	   &	7.6	   &	4.9	   &	3.1	   &	2.1	   &	1.4		\\
14	&	368.5	&	154.9	&	74.6	   &	39.1	   &	22.4	   &	13.3	   &	8.4	   &	5.5	   &	3.6	   &	2.4	   &	1.7		\\
13	&	377.8	&	164.4	&	77.9	   &	41.7	   &	\cellcolor{lightgray}\textbf{24}.\textbf{2}	   &	14.7	   &	9.4	   &	6.3	   &	4.2	   &	2.9	   &	2.0	\\
12      &	379.2	&	170.5	&	82.5	   &	45.2	   &	26.9	   &	16.7	   &	11.0	   &	7.4	   &	5.0	   &	3.5	   &	2.4		\\
11      &	388.9	&	174.6	&	88.6	   &	49.9	   &	30.4	   &	19.3	   &	12.9	   &	8.8	   &	6.1	   &	4.2	   &	3.0		\\
10	&	406.0	&	184.5	&	97.9	   &	57.1	   &	36.0	   &	23.7	   &	16.2	   &	11.3	   &	7.9	   &	5.6	   &	4.0		\\
9	&	440.3	&	214.8	&	113.2  &	68.8	   &	44.8	   &	30.3	   &	21.2	   &	15.0	   &	10.7	   &	7.7	   &	5.5		\\
8	&	470.0	&	237.1	&	138.2  &	87.7	   &	59.6	   &	41.6	   &	29.9	   &	21.8	   &	15.7	   &	11.4	   &	8.1		\\
7	&	563.9	&	307.3	&	181.0  &	120.4  &	86.7	   &	62.1	   &	44.8	   &	31.5	   &	23.3	   &	16.9	   &	12.3		\\
6	&	696.8	&	415.0	&	269.2  &	187.3  &	136.9  &	101.7  &	75.3	   &	57.4	   &	41.8	   &	30.7	   &	23.2		\\
5	&	1007		&	675.0	&	467.8  &	345.8  &	268.0  &	202.3  &	153.3  &	116.9  &	87.3	   &	64.6	   &	47.0	\\
4	&	1839	&	1346	&	997.4  &	765.1  &	586.6  &	451.1  &	349.6  &	266.8  &	200.3  &	147.8  &	109.5	\\
3	&	4800	&	3762	&	2861	   &	2251	   &	1754   &	1358	   &	1041	   &	791.0  &	597.1  &	453.6  &	338.5	\\
  \donescotchrule
  \end{scriptsize}

  \label{tab:reco_cons}
\end{table*}

\section{Expected and observed lower limits on \texorpdfstring{$\Lambda$}{Lambda}}
\label{Limits}

\subsection{Dimuon mass distribution from data}

\begin{figure}[tb]
\begin{center}
\includegraphics[width=\cmsFigWidth]{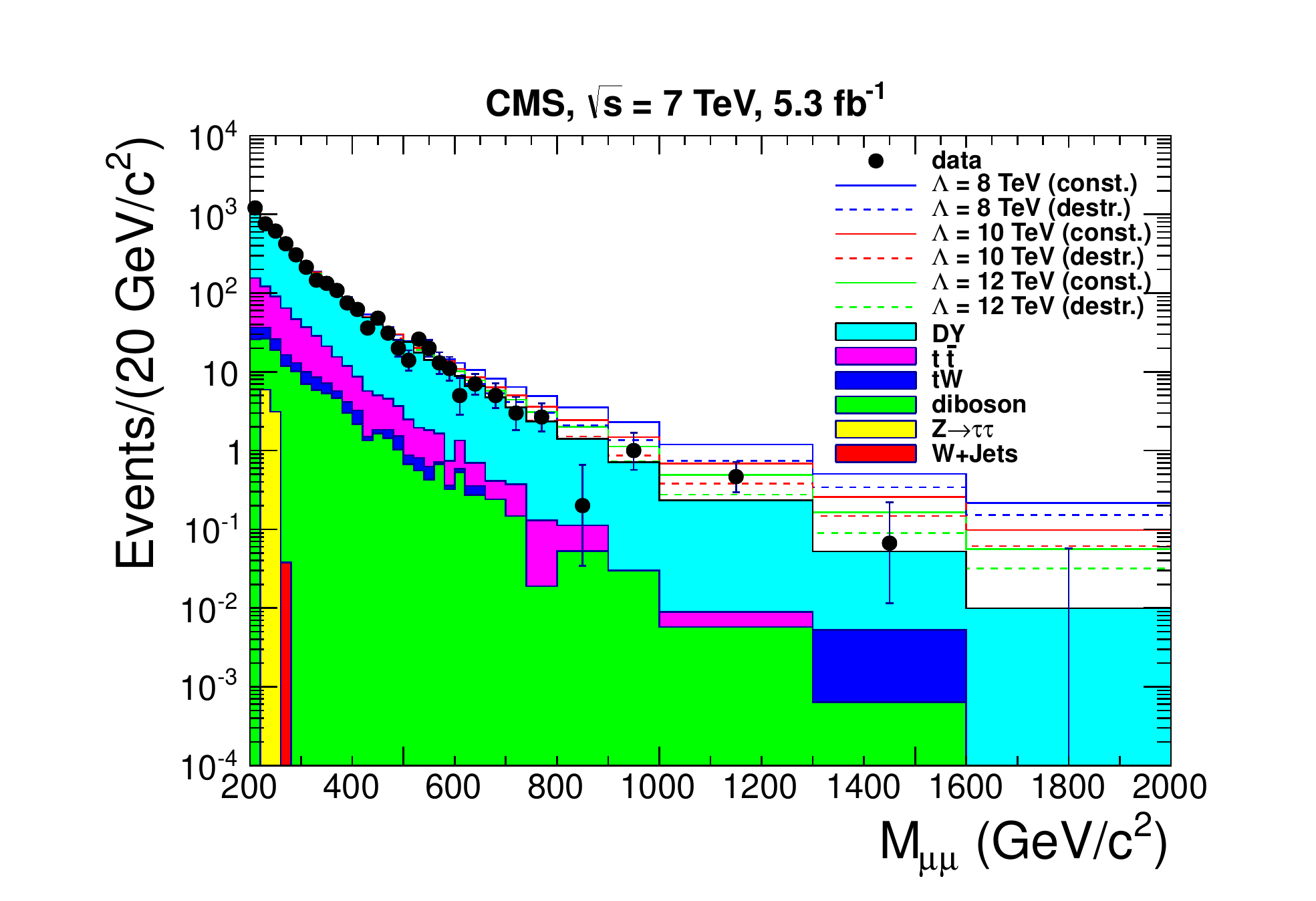}
\end{center}

\caption{Observed spectrum of $M_{\mu\mu}$ and predictions for SM and
  CI/DY plus non-DY SM production.  Predictions are shown for three
  illustrative values of $\Lambda$, for constructive and destructive
  interference.  The error bars for data are 68\% Poisson confidence
  intervals.}

\label{fig:data_mc_comp}
\end{figure}

The observed numbers of events versus $M_{\mu\mu}^{\min}$ are given
in Table \ref{tab:reco_dest}. The observed distribution of
$M_{\mu\mu}$ is plotted in Fig.~\ref{fig:data_mc_comp} along with the
expected distributions from the SM  and for CI/DY plus non-DY SM processes,
for three illustrative
values of $\Lambda$. The data are consistent with
the predictions from the SM, dominated by DY production.

\subsection{Limit-setting procedure}

Since the data are consistent with the SM, we set lower limits on
$\Lambda$ in the context of the LLIM.  The expected and observed 95\%
CL lower limits on $\Lambda$ are determined using the
CL$_\textrm{S}$
modified-frequentist procedure described in ~\cite{READ,JUNK}, taking
the profile likelihood ratio as a test statistic \cite{JUNK2}.  The
expected mean number of events for a signal from CI
is the difference of the number of CI/DY events expected for a given
$\Lambda$, and the number of DY events.  The expected mean number of
background events is the sum of events from the DY process and the
non-DY SM backgrounds.  The observed and expected numbers of events
are given in Tables \ref{tab:reco_dest} and \ref{tab:reco_cons}.

Systematic uncertainties in the predicted signal and background event
yields are estimated from a variety of sources and included as
nuisance parameters in the limit-setting procedure.  Significant
sources of systematic uncertainty are given in Table \ref{tab:sys}.
The uncertainty in the integrated luminosity is described in
Ref.~\cite{LUMI}.  The uncertainty in the CI/DY acceptance is
explained in Section \ref{accept_ci}. The uncertainties in the
prediction of backgrounds depend on the value of $M_{\mu\mu}^{\min}$.
These uncertainties are given in Table \ref{tab:sys} for the
values of $M_{\mu\mu}^{\min}$ chosen for limits on $\Lambda$ with
destructive and constructive interference.  The PDF uncertainty in the
expected yield of DY events is evaluated using the PDF4LHC procedure
~\cite{PDF4LHC}.  The uncertainties in the QED and QCD K-factors are
explained in Sec.~\ref{kfactor}.  The uncertainty from non-DY
backgrounds is due to the statistical uncertainty associated with the
simulated event samples.  The systematic uncertainties which decrease
the limit on $\Lambda$ by the largest amounts are the uncertainties on
the PDF and QED K-factor. When both these uncertainties are set to
zero, the limit for destructive interference is increased by 0.4\% and
the limit for constructive interference is increased by 3.0\%. Thus,
the systematic uncertainties degrade the limits by only small amounts.

We considered possible systematic uncertainties in modeling the
detector response by comparing kinematic distributions between data
and simulation of DY and non-DY SM processes. There are no differences
in these distributions that could lead to significant systematic
uncertainties through their effect on selection efficiency and mass
resolution.

\begin{table} [bt]\centering
  \topcaption{Systematic uncertainties affecting the limit
    on $\Lambda$, evaluated
    for the values of $M_{\mu\mu}^{\min}$
    that provide the best expected limits for
    constructive and destructive interference.
  }
\begin{footnotesize}
\scotchrule[lrr]
        & \multicolumn{2}{c} {Uncertainty (\%) }\\
\cline{2-3} Source & const. & destr. \\
\hline
integrated luminosity & 2.2 & 2.2 \\
acceptance times migration ($A \times M$) & 3.0 & 3.0 \\
PDF & 13.0 & 16.0 \\
QED K-factor & 9.0 & 11.8 \\
QCD K-factor & 3.0 & 3.0 \\
DY MC statistics & 1.2 & 1.6 \\
non-DY backgrounds & 1.1 & 2.9 \\
\donescotchrule
\end{footnotesize}
\label{tab:sys}
\end{table}

\subsection{Results for limits on \texorpdfstring{$\Lambda$}{Lambda}}

\label{sect:results}

\begin{figure}[bt]
\begin{center}
         \includegraphics[width=0.45\textwidth]{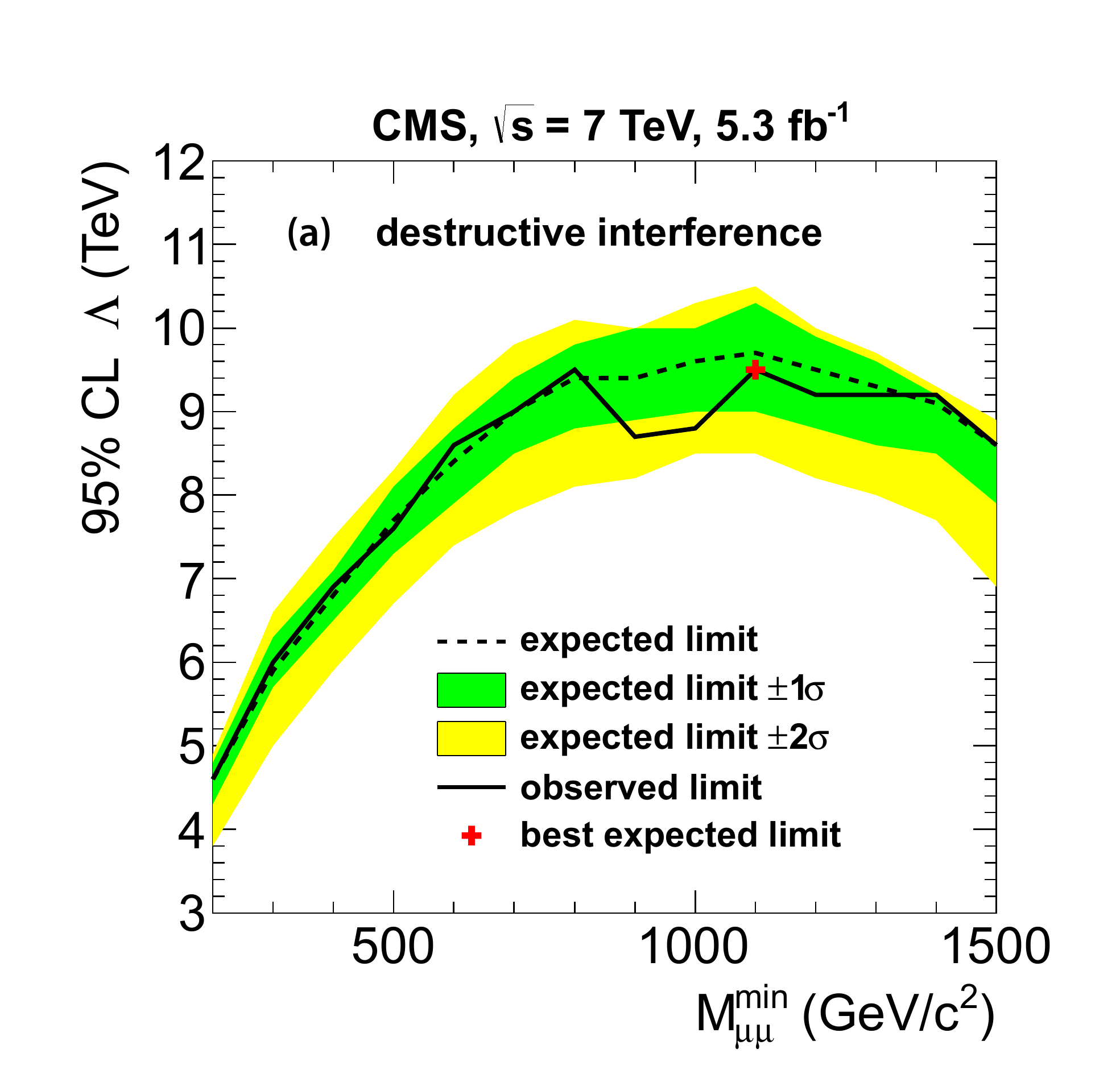}
         \includegraphics[width=0.45\textwidth]{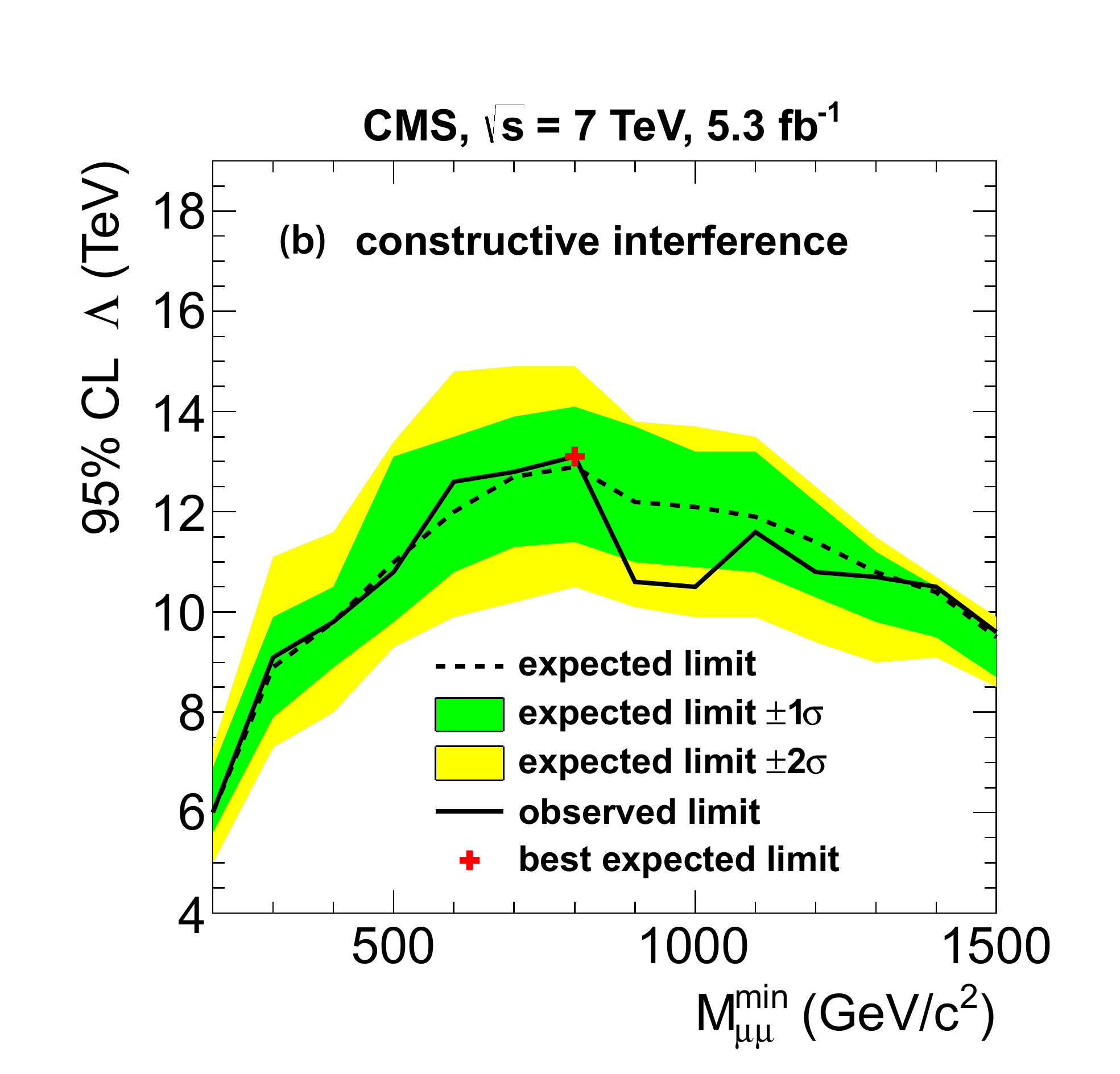}
    \caption{Observed and expected limits as a function of
    $M_{\mu\mu}^{\min}$ for (a) destructive interference and (b)
    constructive interference.  The value of $M_{\mu\mu}^{\min}$,
    chosen to maximize the expected sensitivity, is $1100\GeVcc$ for
    destructive interference, and $800\GeVcc$ for constructive
    interference. The observed (expected) limit is 9.5\TeV (9.7\TeV)
    for destructive interference and 13.1\TeV (12.9\TeV) for
    constructive interference.  The observed limit at the value chosen
    for $M_{\mu\mu}^{\min}$ is indicated with a red plus sign.  The
    variations in the observed limits lie almost entirely within the
    1-$\sigma$ bands, consistent with statistical fluctuations.\label{fig:cl}}

\end{center}
\end{figure}

The observed and expected lower limits on $\Lambda$ at 95\% CL as a
function of $M_{\mu\mu}^{\min}$ for destructive and constructive
interference are shown in Figs.~\ref{fig:cl}(a) and
\ref{fig:cl}(b).  The value of $M_{\mu\mu}^{\min}$, chosen to
maximize the expected sensitivity, is $1100\GeVcc$ for destructive
interference, and $800\GeVcc$ for constructive interference.  The
observed (expected) limit is 9.5\TeV (9.7\TeV) for destructive
interference and 13.1\TeV (12.9\TeV) for constructive interference.
The variations in the observed limits lie almost entirely within the
1-$\sigma$ (standard deviation) uncertainty bands in the expected
limits, consistent with statistical fluctuations.  The number of
expected events corresponding to the observed limits on $\Lambda$ are
highlighted in Tables \ref{tab:reco_dest} and \ref{tab:reco_cons}.

\section{Summary}

The CMS detector is used to measure the invariant mass distribution of
$\mu^+\mu^-$ pairs produced in $\Pp\Pp$ collisions at a center-of-mass
energy of 7\TeV, based on an integrated luminosity of 5.3\fbinv.  The
invariant mass distribution in the range 200 to 2000\GeVcc is found
to be consistent with standard model sources of dimuons, which are
dominated by Drell--Yan production.  The data are interpreted in the
context of a quark- and muon-compositeness model with a left-handed
isoscalar current and an energy scale parameter $\Lambda$. The 95\%
confidence level lower limit on $\Lambda$ is 9.5\TeV under the
assumption of destructive interference between the standard model and
contact-interaction amplitudes.  For constructive interference, the
limit is 13.1\TeV.  These limits are comparable to the most stringent
ones reported to date.

\section*{Acknowledgements}

We congratulate our colleagues in the CERN accelerator departments for
the excellent performance of the LHC and thank the technical and
administrative staffs at CERN and at other CMS institutes for their
contributions to the success of the CMS effort. In addition, we
gratefully acknowledge the computing centres and personnel of the
Worldwide LHC Computing Grid for delivering so effectively the
computing infrastructure essential to our analyses. Finally, we
acknowledge the enduring support for the construction and operation of
the LHC and the CMS detector provided by the following funding
agencies: BMWF and FWF (Austria); FNRS and FWO (Belgium); CNPq, CAPES,
FAPERJ, and FAPESP (Brazil); MEYS (Bulgaria); CERN; CAS, MoST, and
NSFC (China); COLCIENCIAS (Colombia); MSES (Croatia); RPF (Cyprus);
MoER, SF0690030s09 and ERDF (Estonia); Academy of Finland, MEC, and
HIP (Finland); CEA and CNRS/IN2P3 (France); BMBF, DFG, and HGF
(Germany); GSRT (Greece); OTKA and NKTH (Hungary); DAE and DST
(India); IPM (Iran); SFI (Ireland); INFN (Italy); NRF and WCU (Korea);
LAS (Lithuania); CINVESTAV, CONACYT, SEP, and UASLP-FAI (Mexico); MSI
(New Zealand); PAEC (Pakistan); MSHE and NSC (Poland); FCT (Portugal);
JINR (Armenia, Belarus, Georgia, Ukraine, Uzbekistan); MON, RosAtom,
RAS and RFBR (Russia); MSTD (Serbia); SEIDI and CPAN (Spain); Swiss
Funding Agencies (Switzerland); NSC (Taipei); ThEP, IPST and NECTEC
(Thailand); TUBITAK and TAEK (Turkey); NASU (Ukraine); STFC (United
Kingdom); DOE and NSF (USA).

Individuals have received support from the Marie-Curie programme and
the European Research Council (European Union); the Leventis
Foundation; the A. P. Sloan Foundation; the Alexander von Humboldt
Foundation; the Belgian Federal Science Policy Office; the Fonds pour
la Formation \`a la Recherche dans l'Industrie et dans l'Agriculture
(FRIA-Belgium); the Agentschap voor Innovatie door Wetenschap en
Technologie (IWT-Belgium); the Ministry of Education, Youth and Sports
(MEYS) of Czech Republic; the Council of Science and Industrial
Research, India; the Compagnia di San Paolo (Torino); and the HOMING
PLUS programme of Foundation for Polish Science, cofinanced from
European Union, Regional Development Fund.

\bibliography{auto_generated}   
\cleardoublepage \appendix\section{The CMS Collaboration \label{app:collab}}\begin{sloppypar}\hyphenpenalty=5000\widowpenalty=500\clubpenalty=5000\textbf{Yerevan Physics Institute,  Yerevan,  Armenia}\\*[0pt]
S.~Chatrchyan, V.~Khachatryan, A.M.~Sirunyan, A.~Tumasyan
\vskip\cmsinstskip
\textbf{Institut f\"{u}r Hochenergiephysik der OeAW,  Wien,  Austria}\\*[0pt]
W.~Adam, E.~Aguilo, T.~Bergauer, M.~Dragicevic, J.~Er\"{o}, C.~Fabjan\cmsAuthorMark{1}, M.~Friedl, R.~Fr\"{u}hwirth\cmsAuthorMark{1}, V.M.~Ghete, J.~Hammer, N.~H\"{o}rmann, J.~Hrubec, M.~Jeitler\cmsAuthorMark{1}, W.~Kiesenhofer, V.~Kn\"{u}nz, M.~Krammer\cmsAuthorMark{1}, I.~Kr\"{a}tschmer, D.~Liko, I.~Mikulec, M.~Pernicka$^{\textrm{\dag}}$, B.~Rahbaran, C.~Rohringer, H.~Rohringer, R.~Sch\"{o}fbeck, J.~Strauss, A.~Taurok, W.~Waltenberger, G.~Walzel, E.~Widl, C.-E.~Wulz\cmsAuthorMark{1}
% ...
\vskip\cmsinstskip
\dag:~Deceased\\
1:~~Also at Vienna University of Technology, Vienna, Austria\\
% ...

\end{sloppypar}
\end{document}